\documentclass[aps, twocolumn]{revtex4}

\usepackage{graphicx}
\usepackage{amsmath}
\usepackage{dsfont}
\usepackage{amssymb}
\usepackage{physics}
\usepackage{hyperref}
\usepackage{ulem}
\usepackage{natbib}
\usepackage{float}
\hypersetup{colorlinks=true,
	    final=true,
	    linkcolor=blue,
	    citecolor=blue,
	    filecolor=blue,
	    urlcolor=blue,}

\usepackage{amssymb}
\usepackage{mathrsfs}

\newcommand{\beq}{\begin{equation}}
\newcommand{\eeq}{\end{equation}}
\newcommand{\ba}{\begin{align}}
\newcommand{\ea}{\end{align}}

\newcommand{\bn}{\begin{equation*}}
\newcommand{\en}{\end{equation*}}

\begin{document}

\title{Statistical Mechanics of Paraparticles}

\author{Nupoor Thakur and Navinder Singh}

\email{nupoorthakur5@gmail.com; navinder.phy@gmail.com; Phone: +91 9662680605}
\affiliation{Department of Physics \& Photonics Science, N.I.T. Hamirpur, Himachal Pradesh, India. PIN: 177005. Theoretical Physics Division, Physical Research Laboratory (PRL), Ahmedabad, India. PIN: 380009.}

\begin{abstract}
Quantum mechanics broadly classifies the particles into two categories: $(1)$ fermions and $(2)$ bosons. Fermions are half-integer spin particles, obeying Pauli's exclusion principle and Fermi-Dirac statistics. Whereas bosons are integer spin particles, not obeying Pauli's exclusion principle, and obeying Bose-Einstein statistics. However, there are two exceptions to this standard case: first, anyons, which exist in 2-dimensional systems, and secondly, paraparticles, which can exist in any dimension. Paraparticles follow their non-trivial parastatistics, obeying their generalised exclusion principle. In this paper, we provide a detailed review of the foundations of paraparticle statistics established in \cite{wang2025particle}. We extend this work further and then derive an important expression for the heat capacity of paraparticles for a specific case, which would provide a handle for the experimental detection of paraparticles in appropriate systems. 
\end{abstract}

\maketitle

 \section{Introduction} 
 
 In quantum mechanics, a many-particle state is defined by the wave function $\psi(x_1, x_2, ...,x_n)$ which depends on the position coordinates of the particles. When we swap two particles, it does not change the physical state, but it does change the phase of the wavefunction. If we swap a particle at position $x_1$ with a particle at position $x_2$ we get:
\beq
\psi(x_2, x_1, ...,x_n) = c\psi(x_1, x_2, ...,x_n)
\eeq
If we swap the same two particles again, then the system should go back to its original state.
Performing a second exchange:
\begin{equation}
\begin{aligned}
\psi(x_1, x_2, ..., x_n) &= c\psi(x_2, x_1, ...,x_n) \\
                                 &=c^2\psi(x_1, x_2, ...,x_n)\
\end{aligned}
\end{equation}
 Since the wavefunction must remain non-zero \( \psi \neq 0 \), this implies \( c^2=1  \), which gives us two solutions: (1) $c = +1:$ The wavefunction does not change sign when the particles are exchanged. These particles are called bosons. (2) $c = -1:$ The wavefunction acquires a negative sign when particles are exchanged. These particles are called fermions.

Nevertheless, there are two notable exceptions to these quantum exchange statistics: the first is anyons, which can be found solely in two dimensions, and the second is paraparticles that adhere to parastatistics, which can exist in any dimensional space \cite{wang2025particle}.

 For anyons (in 2D) $\psi(x_2, x_1, ..., x_n) =e^{i\theta}\psi(x_1, x_2, ..., x_n)$, where \(e^{i\theta} =\) phase factor and \(\theta\) can take the value from $0$ to \(\pi \).

However, the exchange rules for paraparticles are as follows: the wavefunction for paraparticles is written as $\psi^I(\{x_i\}_{i=1}^n)$ where superscript $I$ is a collective index denoting all the flavours (internal quantum numbers) of paraparticles. $x_1, x_2, ..., x_n$ denotes the modes of paraparticles. Modes can be position coordinates, momentum states, etc.

  The most fundamental aspect is the exchange rule. When two paraparticles $(x_j)$ and $(x_{j+1})$ are swapped, the exchange rule is given as:

\beq
\psi^I(\{x_i\}_{i=1}^n)|_{x_j \leftrightarrow x_{j+1}}= \sum_J(R)_{J}^I \psi^J (\{x_i\}_{i=1}^n)
\eeq

\begin{description}
\item \(\sum_J \): Summation over all the possible flavours (internal quantum states) labelled by $J$ and \(|_{x_j \leftrightarrow x_{j+1}} \) means swap particles $j$ and $j+1$ (exchange their modes).
\item \((R)_{J}^I\): matrix associated with the exchange operation which reshuffles flavours (internal states) instead of just applying a simple phase change. This is where the paraparticle exchange statistics differ from those of fermions and bosons.

\end{description}


\section{Commutation Relations for $R$-Matrix}

The creation and annihilation operators for paraparticles are defined as:
\begin{description}
\item \(\hat{\psi}_{i,a}^-\) : Annihilation Operator - Removes a paraparticle in mode $i$ of flavour $a$ (internal quantum number).
\item \(\hat{\psi}_{j,b}^+\) : Creation Operator - Adds a paraparticle in mode $j$ of flavour $b$.
\end{description}

For a given $R$-matrix, the paraparticle creation and annihilation operators are defined through the commutation relations \cite{wang2025particle}:

\beq
\hat{\psi}_{i,a}^- \hat{\psi}_{j,b}^+ = \sum_{c,d} R^{ac}_{bd} \, \hat{\psi}_{j,c}^+ \hat{\psi}_{i,d}^- +  \delta_{ab}\delta_{ij}  
\eeq

\beq
\hat{\psi}_{i,a}^+ \hat{\psi}_{j,b}^+ = \sum_{c,d} R_{ab}^{cd} \, \hat{\psi}_{j,c}^+ \hat{\psi}_{i,d}^+ 
\eeq

\beq
\hat{\psi}_{i,a}^- \hat{\psi}_{j,b}^- = \sum_{c,d} R_{dc}^{ba} \, \hat{\psi}_{j,c}^- \hat{\psi}_{i,d}^- 
\eeq
where $i$ and $j$ are different modes (for example, position coordinates or momentum states) and $a, b, c$ and $d$ are the flavours (for example, spin, etc).

\subsection{Contracted Bilinear Operator}

A contracted bilinear operator tracks how one type of paraparticle transforms into another, or it allows us to describe how paraparticles move between different modes while keeping track of their flavours. It is defined as:

\beq
\hat{e}_{ij} = \sum_{a=1}^{{\mathscr{M}}} \hat{\psi}_{i,a}^+ \hat{\psi}_{j,a}^-
\eeq


 This operator obeys Lie algebra rules \cite{wang2025particle}. The commutative relations of this operator with creation and annihilation operators, as well as with itself, tell us about the transformation of flavours. In the following, we have derived these expressions in detail:

\subsubsection{To prove: $[\hat{e}_{ij}, \hat{\psi}_{k,b}^+] = \delta_{jk} \hat{\psi}_{i,b}^+$}

 Proof:

\begin{equation*}
[\hat{e}_{ij}, \hat{\psi}_{k,b}^+] = \sum_{a=1}^{\mathscr{M}} [\hat{\psi}_{i,a}^+ \hat{\psi}_{j,a}^-, \hat{\psi}_{k,b}^+] 
\end{equation*}

\begin{equation*}
 = \sum_a [\hat{\psi}_{i,a}^+ , \hat{\psi}_{k,b}^+] \hat{\psi}_{j,a}^- + \hat{\psi}_{i,a}^+ [\hat{\psi}_{j,a}^- , \hat{\psi}_{k,b}^+] 
\end{equation*}

\begin{multline*}
= \sum_a  ({\hat{\psi}_{i,a}^+\hat{\psi}_{k,b}^+\hat{\psi}_{j,a}^-} -{\hat{\psi}_{k,b}^+\hat{\psi}_{i,a}^+\hat{\psi}_{j,a}^-} \\
+{\hat{\psi}_{i,a}^+\hat{\psi}_{j,a}^-\hat{\psi}_{k,b}^+}-{\hat{\psi}_{i,a}^+\hat{\psi}_{k,b}^+\hat{\psi}_{j,a}^-} )
\end{multline*}

\begin{equation*}
= \sum_a \left({\hat{\psi}_{i,a}^+\hat{\psi}_{j,a}^-\hat{\psi}_{k,b}^+}-{\hat{\psi}_{k,b}^+\hat{\psi}_{i,a}^+\hat{\psi}_{j,a}^-} \right)
\end{equation*}

The product ${\hat{\psi}_{j,a}^-\hat{\psi}_{k,b}^+}$ is replaced using equation $(4)$, we get:

\begin{equation*}
= \sum_a \hat{\psi}_{i,a}^+ \left(\sum_{c,d} R_{bd}^{ac} \hat{\psi}_{k,c}^+ \hat{\psi}_{j,d}^-  + \delta_{jk} \delta_{ab}\right) - \sum_a \hat{\psi}_{k,b}^+ \hat{\psi}_{i,a}^+ \hat{\psi}_{j,a}^- 
\end{equation*}

\begin{equation*}
= \sum_{a,c,d} R_{bd}^{ac} \hat{\psi}_{i,a}^+ \hat{\psi}_{k,c}^+ \hat{\psi}_{j,d}^- + \sum_a \hat{\psi}_{i,a}^+ \delta_{jk} \delta_{ab} - \sum_a \hat{\psi}_{k,b}^+ \hat{\psi}_{i,a}^+ \hat{\psi}_{j,a}^- 
\end{equation*}
\begin{equation*}
=  \sum_d \left(\sum_{a,c}R_{bd}^{ac} \hat{\psi}_{i,a}^+ \hat{\psi}_{k,c}^+\right) \hat{\psi}_{j,d}^- + \delta_{jk} \hat{\psi}_{i,b}^+ - \sum_a \hat{\psi}_{k,b}^+ \hat{\psi}_{i,a}^+ \hat{\psi}_{j,a}^-
\end{equation*}

$ \left(\sum_{a,c}R_{bd}^{ac} \hat{\psi}_{i,a}^+ \hat{\psi}_{k,c}^+\right)$ is replaced using equation (5) and we get:

\[
 = \sum_{d} \hat{\psi}_{k,b}^+ \hat{\psi}_{i,d}^+ \hat{\psi}_{j,d}^- + \delta_{jk} \hat{\psi}_{i,b}^+ - \sum_a \hat{\psi}_{k,b}^+ \hat{\psi}_{i,a}^+ \hat{\psi}_{j,a}^-
\]
Using equation (7), $\sum_{d} \hat{\psi}_{k,b}^+ \hat{\psi}_{i,d}^+ \hat{\psi}_{j,d}^-$ becomes $ \hat{\psi}_{k,b}^+\hat{e}_{ij}$ and $\sum_a \hat{\psi}_{k,b}^+ \hat{\psi}_{i,a}^+ \hat{\psi}_{j,a}^-$ becomes $\hat{\psi}_{k,b}^+ \hat{e}_{ij}$ and the two terms cancel each other and we get: 

\beq
[\hat{e}_{ij}, \hat{\psi}_{k,b}^+] = \delta_{jk} \hat{\psi}_{i,b}^+
\eeq

\subsubsection{To prove: $[\hat{e}_{ij}, \hat{\psi}_{k,b}^-] = -\delta_{ki} \hat{\psi}_{j,b}^-$}

Proof:

\bn
[\hat{e}_{ij}, \hat{\psi}_{k,b}^-] = \sum_{a=1}^{\mathscr{M}} [\hat{\psi}_{i,a}^+ \hat{\psi}_{j,a}^-, \hat{\psi}_{k,b}^-] 
\en
\bn
= \sum_a [\hat{\psi}_{i,a}^+ , \hat{\psi}_{k,b}^-] \hat{\psi}_{j,a}^- + \hat{\psi}_{i,a}^+ [\hat{\psi}_{j,a}^- , \hat{\psi}_{k,b}^-] 
\en

\begin{multline*}
= \sum_a ({\hat{\psi}_{i,a}^+\hat{\psi}_{k,b}^-\hat{\psi}_{j,a}^-} -{\hat{\psi}_{k,b}^-\hat{\psi}_{i,a}^+\hat{\psi}_{j,a}^-} \\
+{\hat{\psi}_{i,a}^+\hat{\psi}_{j,a}^-\hat{\psi}_{k,b}^-}-{\hat{\psi}_{i,a}^+\hat{\psi}_{k,b}^-\hat{\psi}_{j,a}^-} )
\end{multline*}

\bn
= \sum_a  \left({-{\hat{\psi}_{k,b}^-\hat{\psi}_{i,a}^+\hat{\psi}_{j,a}^-}+\hat{\psi}_{i,a}^+\hat{\psi}_{j,a}^-\hat{\psi}_{k,b}^-} \right)
\en

The product $\hat{\psi}_{k,b}^-\hat{\psi}_{i,a}^+$ is replaced using equation $(4)$, we get:

\bn
 =-\sum_{a}  \left(\sum_{c,d}R_{ad}^{bc} \hat{\psi}_{i,c}^+ \hat{\psi}_{k,d}^- + \delta_{ba}\delta_{ki}\right)  \hat{\psi}_{j,a}^- + \sum_a \hat{\psi}_{i,a}^+ \hat{\psi}_{j,a}^- \hat{\psi}_{k,b}^- 
\en

\bn
= -\sum_{a,c,d} R_{ad}^{bc} \hat{\psi}_{i,c}^+ \hat{\psi}_{k,d}^- \hat{\psi}_{j,a}^- - \sum_a \hat{\psi}_{j,a}^- \delta_{ba}\delta_{ki} + \sum_a \hat{\psi}_{i,a}^+ \hat{\psi}_{j,a}^- \hat{\psi}_{k,b}^- 
\en
\bn
= -\sum_{c} \hat{\psi}_{i,c}^+ \left(\sum_{a,d}R_{ad}^{bc} \hat{\psi}_{k,d}^- \hat{\psi}_{j,a}^-\right) - \delta_{ki} \hat{\psi}_{j,b}^-  + \sum_a \hat{\psi}_{i,a}^+ \hat{\psi}_{j,a}^- \hat{\psi}_{k,b}^-
\en

$\left(\sum_{a,d}R_{ad}^{bc} \hat{\psi}_{k,d}^- \hat{\psi}_{j,a}^-\right)$ is replaced using equation (6) and we get:

\bn = -\sum_{c} \hat{\psi}_{i,c}^+ \hat{\psi}_{j,c}^- \hat{\psi}_{k,b}^- - \delta_{ki} \hat{\psi}_{j,b}^-  + \sum_a \hat{\psi}_{i,a}^+ \hat{\psi}_{j,a}^- \hat{\psi}_{k,b}^-
\en

Using equation (7), $\sum_{c} \hat{\psi}_{i,c}^+ \hat{\psi}_{j,c}^- \hat{\psi}_{k,b}^-$ becomes $ \hat{e}_{ij} \hat{\psi}_{k,b}^-$ and $\sum_a \hat{\psi}_{i,a}^+ \hat{\psi}_{j,a}^- \hat{\psi}_{k,b}^-$ becomes $ \hat{e}_{ij} \hat{\psi}_{k,b}^-$ and the two terms cancel each other and we get:

\beq
[\hat{e}_{ij}, \hat{\psi}_{k,b}^-] = -\delta_{ki} \hat{\psi}_{j,b}^-
\eeq

\subsubsection{To prove: $[\hat{e}_{ij}, \hat{e}_{kl}] =\delta_{jk} \hat{e}_{il} - \delta_{il} \hat{e}_{kj}$}

Proof:

\begin{align*}
[\hat{e}_{ij}, \hat{e}_{kl}] &= \sum_b [\hat{e}_{ij}, \hat{\psi}_{k,b}^+\hat{\psi}_{l,b}^-] \\
                                      &=  \sum_b [\hat{e}_{ij}, \hat{\psi}_{k,b}^+]\hat{\psi}_{l,b}^- + \sum_b \hat{\psi}_{k,b}^+[\hat{e}_{ij}, \hat{\psi}_{l,b}^-]
\end{align*}
Using equation $(8)$ and $(9)$:

\[
[\hat{e}_{ij}, \hat{e}_{kl}] = \sum_b \delta_{jk} \hat{\psi}_{i,b}^+\hat{\psi}_{l,b}^- - \sum_b \hat{\psi}_{k,b}^+\delta_{il} \hat{\psi}_{j,b}^-
\]

\beq
[\hat{e}_{ij}, \hat{e}_{kl}] =\delta_{jk} \hat{e}_{il} - \delta_{il} \hat{e}_{kj}
\eeq

\subsection{Particle Number Operator}
This operator counts the number of particles in a particular mode, summing over flavours. Mathematically:
\[
\hat{n}_i = \hat{e}_{ii} = \sum_{a=1}^{\mathscr{M}}  \hat{\psi}_{i,a}^+\hat{\psi}_{i,a}^-
\]
The commutation relations for the paraparticle number operator are as follows:
\begin{enumerate}
 \item $[\hat{n}_i, \hat{n}_j]  = [\hat{e}_{ii}, \hat{e}_{jj}] = \delta_{ij} \hat{e}_{ij} - \delta_{ij} \hat{e}_{ji}=0$
\item $[\hat{n}_i, \hat{\psi}_{j,b}^+]= [\hat{e}_{ii}, \hat{\psi}_{j,b}^+] = \delta_{ij} \hat{\psi}_{i,b}^+$
\item $[\hat{n}_i, \hat{\psi}_{j,b}^-]= [\hat{e}_{ii}, \hat{\psi}_{j,b}^-] = -\delta_{ij} \hat{\psi}_{i,b}^-$
Combining the two: $[\hat{n}_i, \hat{\psi}_{j,b}^{\pm}]=\pm \delta_{ij} \hat{\psi}_{j,b}^{\pm} $
\item Total number operator: $\hat{n}=\sum_i \hat{n}_i$ , combining this with the second relation we get: $[ \hat{n}, \hat{\psi}_{j,b}^+]=\hat{\psi}_{j,b}^+$

\end{enumerate}

\subsection{Mathematical Proof of the Exclusion Principle for Paraparticles (Para-exclusion Principle)}
For one of the paraparticle statistics cases where there can be only one paraparticle in a given mode we have $\hat{\psi}_{i,a}^+ \hat{\psi}_{i,b}^+ =0$ which can be proved by using the specific R statistics (Table 1 in \cite{wang2025particle}):

Taking the value of $R_{cd}^{ab} $ from Table 1 (Example 3), $R_{cd}^{ab} = -\delta_{ac}\delta_{bd}$, and putting it in equation (5), we get:

\[
\hat{\psi}_{i,a}^+ \hat{\psi}_{j,b}^+ = - \sum_{c,d}\delta_{ac}\delta_{bd}\hat{\psi}_{j,c}^+\hat{\psi}_{i,d}^+
\]

\beq
\hat{\psi}_{i,a}^+ \hat{\psi}_{j,b}^+ = -\hat{\psi}_{j,a}^+\hat{\psi}_{i,b}^+
\eeq

Taking $i=j$ (same mode), $\hat{\psi}_{i,a}^+ \hat{\psi}_{i,b}^+ = -\hat{\psi}_{i,a}^+\hat{\psi}_{i,b}^+$
or $2\hat{\psi}_{i,a}^+ \hat{\psi}_{i,b}^+ =0$ $\Rightarrow$ $\hat{\psi}_{i,a}^+ \hat{\psi}_{i,b}^+ =0$. This means that a single mode (the above formula has used mode $i$) can not be occupied by two paraparticles even if they have different flavours $(a{\neq}b)$.

\subsection{State Space}

The construction of the state space of the system containing multiple paraparticles is done by applying the creation operator on the vacuum (just like in the case of fermion Fock Space) \cite{wang2025particle}.
\beq
|\psi\rangle = \hat{\psi}_{i_1,a_1}^+ \hat{\psi}_{i_2,a_2}^+ \cdots \hat{\psi}_{i_n,a_n}^+ |0\rangle
\eeq

where, $n$ is the total number of paraparticles, $i$ represents the different modes and $a$ represents the different flavours.

\subsection{Unitary Exchange Operator}

The initial quantum state refers to a situation where we have two paraparticles located at different positions in space, mathematically: $|0:ia,jb\rangle=\hat{\psi}_{i,a}^+ \hat{\psi}_{j,b}^+|0\rangle$, where, $|0\rangle$ means vacuum state, no particles yet and \(\hat{\psi}_{i,a}^+\) is defined in starting of section $\textbf{II}$. The Unitary Exchange Operator $\hat{E}_{ij}$ is defined such that it swaps the position of paraparticles at modes $i$ and $j$ \cite{wang2025particle}.
\begin{equation}
\begin{aligned}
 \hat{E}_{ij}\hat{\psi}_{i,a}^+\hat{E}_{ij}^\dagger=\hat{\psi}_{j,a}^+ \\
\hat{E}_{ij}\hat{\psi}_{j,a}^+\hat{E}_{ij}^\dagger=\hat{\psi}_{i,a}^+ 
\end{aligned}
\end{equation}

which moves particle $i$ to $j$ and vice versa.

 If we apply $\hat{E}_{ij}$ to original state:
\bn
\hat{E}_{ij} \, |0; ia, jb\rangle = \hat{E}_{ij} \hat{\psi}_{i,a}^+ \hat{\psi}_{j,b}^+|0\rangle
\en
\bn
= \left( \hat{E}_{ij} \hat{\psi}_{i,a}^+ \hat{E}_{ij}^{\dagger} \right) \left( \hat{E}_{ij} \hat{\psi}_{j,b}^+ \hat{E}_{ij}^{\dagger} \right)\hat{E}_{ij} |0\rangle
\en

Using equations $(13)$, and vacuum remains invariant under any operator. Also using equation $(5)$ in next step to change $\hat{\psi}_{j,a}^+ \hat{\psi}_{i,b}^+$ to $\sum_{a',b'} R^{b'a'}_{ab} \hat{\psi}_{i,b'}^+ \hat{\psi}_{j,a'}^+$.
\bn
                                             = \hat{\psi}_{j,a}^+ \hat{\psi}_{i,b}^+ |0\rangle = \sum_{a',b'} R^{b'a'}_{ab}  \hat{\psi}_{i,b'}^+ \hat{\psi}_{j,a'}^+  |0\rangle
\en
\bn
                                         \hat{E}_{ij} \, |0; ia, jb\rangle    = \sum_{a',b'} R^{b'a'}_{ab} \, |0; i b', j a'\rangle 
\en

Paraparticles swap their positions, and the flavours transform according to the R-matrix.
The unitary rotation shows that paraparticles do not simply swap like fermions or bosons, but their states transform non-trivially.

\section{Exact Solution of a Free Paraparticle}

Hamiltonian $\hat{H}$ describes the total energy of the system of free (non-interacting) paraparticles \cite{wang2025particle}:

\begin{equation}
\begin{aligned}
\hat{H} = \sum_{1 \leq i,\, j \leq N} h_{ij} \, \hat{e}_{ij} = \sum_{1 \leq i,\, j \leq N} \sum_{1 \leq a \leq {\mathscr{M}}} h_{ij} \, \hat{\psi}_{i,a}^+ \hat{\psi}_{j,a}^-
\end{aligned}
\end{equation}

where, \(i, j = 1, 2, ..., N\) are the modes (could be lattice sites or momentum states), \(a = 1, 2, ..., {\mathscr{M}}\) are the flavours (internal states), \(h_{ij}\) is the interaction strength between modes $i$ and $j$ and \(h_{ij}=h_{ij}^*\) is Hermitian matrix that encodes how paraparticles move between the modes.

We need to diagonalise the Hamiltonian for further computation.

  The current form is complicated because paraparticles at different sites are mixed through $h_{ij}$ - the Hamiltonian has off-diagonal terms in $i$, $j$ where modes are coupled. This makes time evolution and observables difficult to compute. We want to diagonalise the Hamiltonian, that is, find a new set of operators where it is decoupled so that separate modes have separate Hamiltonians.

Defining new operators \(\hat{\psi}_{i,a}^-=\sum_{k=1}^N U_{ki}^* \tilde{\psi}_{k,a}^- \) and \(\hat{\psi}_{i,a}^+=\sum_{k=1}^N U_{ki} \tilde{\psi}_{k,a}^+ \), where $U$\(\in{C}^{N \times N}\)=  a unitary matrix that diagonalizes $h$. \( U h U^\dagger={diag}({\epsilon}_1, ..., {\epsilon}_N) \) and \(U U^\dagger=U^\dagger U=1 \). Matrix h is diagonalised as \(h_{kp}=(U h U^\dagger)_{kp}={\epsilon}_k \delta_{kp} \) or \(\sum_{ij} U_{ki} h_{ij} U_{pj}^*={\epsilon}_k \delta_{kp} \), which gives us \({\epsilon}_k = \sum_{ij} U_{ki} h_{ij} U_{kj}^*\).

After the Transformation, the Hamiltonian becomes

\beq
\hat{H} = \sum_{k=1}^N {\epsilon}_k  \tilde{n}_k
= \sum_{k=1}^N {\epsilon}_k \sum_{a=1}^{\mathscr{M}} \tilde{\psi}_{k,a}^+ \tilde{\psi}_{k,a}^-
\eeq

where, \(\epsilon_k =\) eigenvalues of $h_{ij}$. 

  These are the energy levels of the system, each corresponding to a decoupled mode labelled by $k$. Physically, they describe the energy required to occupy the mode $k$. Now $\hat{H}$ has a simple form; each paraparticle has its energy level. The system now behaves like a collection of independent modes. Each mode is characterised by energy $\epsilon$ and can be occupied by a paraparticle of different flavours ${\mathscr{M}}$.


\section{Grand Canonical Partition Function}

The partition function $Z$ of the whole system of paraparticles is a product of single-mode partition functions $z_R (x_k)$ \cite{wang2025particle}.
\beq
Z(\beta)=Tr[e^{-\beta \hat{H}}] = \prod_k z_R (x_k)
\eeq
where, \(x_k=e^{-\beta (\epsilon_k-\mu)}\) is the Boltzmann Factor, \(\beta=\frac{1}{k_b T}\), \(\epsilon_k\) is the energy of mode $k$ and $\mu$ is the chemical potential of the system.

 The partition function $Z$ encodes all the thermodynamic information. The trace is over the State space of paraparticles. In the case of free paraparticles, the Hamiltonian has been diagonalised into modes, and each mode contributes independently to the partition function $Z$.


Single-mode partition function is given as \cite{wang2025particle}:
\beq
 z_k(x_k) = Tr[e^{-\beta \epsilon \hat{n}}]  = \sum_{n=0}^{n_{max}} d_n x_k^n 
\eeq
where $n$ is the number of paraparticles occupying the mode and \textbf {$d_n$ are non-negative integers that define the generalised exclusion statistics for the paraparticles associated with $R$.} For bosons, there is no exclusion, so \( d_n = 1 \) for all \( n \geq 0 \). For fermions, the Pauli exclusion principle allows at most one particle per mode, hence \( d_0 = d_1 = 1 \), and \( d_n = 0 \) for \( n \geq 2 \). Mode $k$ defines one particle states. It is different from the wave number $k$, which can accommodate two electrons of opposite spin. From now onwards, $k$ is defined as one particle state; otherwise, it will be explicitly mentioned if it is wavevector $k$, so that confusion should not arise. For paraparticles, the allowed occupation numbers are governed by a specific \( R \)-matrix, leading to a generalised exclusion rule where \( d_n \) depends on both \( n \) and flavours (internal degrees of freedom). For example for \( R \)-matrix defined by exapmle 3 in the table (Table 1 below) we have \( d_0 = 1, d_1 = {\mathscr{M}} \), and \( d_n = 0 \) for \( n \geq 2 \), where \( {\mathscr{M}} \) is the number of flavours of the paraparticles. For \( R \)-matrix defined by example 4 in the table (Table 1 below), we have \( d_0 = 1, d_1 = {\mathscr{M}}, d_2 = 1 \), and \( d_n = 0 \) for \( n \geq 3 \).

 The partition function can be derived as: The Hamiltonian for a system of non-interacting paraparticles is given by: $\hat{H} = \sum_k \epsilon_k \hat{n}_k$, where $\epsilon_k$ is the energy of mode $k$, and $\hat{n}_k$ is the number operator for mode $k$. For the ease of calculations, we have absorbed the chemical potential $\mu$ into $\epsilon_k$. The canonical partition function is $Z = \text{Tr}[e^{-\beta \hat{H}}] = \text{Tr}[e^{-\beta \sum_k \epsilon_k \hat{n}_k}]$. Because the number operators commute \( ([\hat{n}_k, \hat{n}_{k'}] = 0) \), the exponential factors into a product $e^{-\beta \sum_k \epsilon_k \hat{n}_k} = \prod_k e^{-\beta \epsilon_k \hat{n}_k}$. The trace of the product becomes a product of traces over independent modes $Z = \text{Tr} \left[ \prod_k e^{-\beta \epsilon_k \hat{n}_k} \right] = \prod_k \text{Tr} [ e^{-\beta \epsilon_k \hat{n}_k} ]$. The single-mode partition function is given as $z_k(x_k) = \sum_{n=0}^\infty d_n e^{-\beta \epsilon_k n}= \text{Tr} [ e^{-\beta \epsilon_k \hat{n}_k} ]$. Therefore, the partition function $Z$ can be written as a product of the single-mode partition function. Hence proved $Z=Tr[e^{-\beta \hat{H}}] = \prod_k z_R (x_k)$.


\section{Average Occupation Number }

We derive the average occupation number $\langle \tilde{n}_k \rangle$ for a paraparticle defined by $R$ statistics of Example 3 and Example 4 of Table 1 in reference \cite{wang2025particle}, which we redraw for completeness here. 
\begin{table}[h]
\centering
\caption{Examples of \( R \) matrices and their single-mode partition functions \( z_R(x) \)}
\label{tab:R_matrices}
\begin{tabular}{|c|c|c|c|c|}
\hline
\textbf{Ex.} & \textbf{1} & \textbf{2} & \textbf{3} & \textbf{4} \\ \hline
\( R_{cd}^{ab} \) & \( -\delta_{ad}\delta_{bc} \) & \( \delta_{ad}\delta_{bc}(-1)^{\delta_{ab}} \) & \( -\delta_{ac}\delta_{bd} \) & \( \lambda_{ab}\epsilon_{cd}-\delta_{ac}\delta_{bd} \) \\ \hline
\( z_R(x) \) & \( (1+x)^{\mathscr{M}} \) & \( (1+x)^{\mathscr{M}} \) & \( 1+{\mathscr{M}}x \) & \( 1+{\mathscr{M}}x+x^2 \) \\ \hline
\end{tabular}
\end{table}

In Example $1$, the $R$-matrix is given by $-\delta_{ad}\delta_{bc} $. This rule behaves similarly to fermions but for multiple flavours (internal quantum numbers) of fermions. In this system, there are ${\mathscr{M}}$ flavours of fermions, each one is allowed to occupy the mode (one particle state) individually, but no two particles of the same flavour can exist in the same mode. The partition function in this example is given by $(1+x)^{\mathscr{M}} $. In Example $2$, the $R$-matrix is given by $ \delta_{ad}\delta_{bc}(-1)^{\delta_{ab}} $. In this case, the behaviour is identical to ${\mathscr{M}}$ flavour fermions if and only if $a=b$. Otherwise, it exhibits non-trivial statistics. However, the same exclusion rule is enforced such that only one particle of any one of the flavour ${\mathscr{M}}$ is permitted in each mode. The partition function in this example is given by $(1+x)^{\mathscr{M}} $. Example $3$ is the most interesting one, the $R$-matrix is given by $-\delta_{ac}\delta_{bd} $. In this case, even when paraparticles are of different flavours, the occupation of the same mode by two paraparticles is prohibited. This represents a stricter condition than fermions. The partition function in this example is given by $(1+{\mathscr{M}}x) $. For each mode, two possibilities are permitted: (1)  The mode may remain unoccupied (1 configuration). (2) The mode may be occupied by a single paraparticle of any type (${\mathscr{M}}$ possible configurations). This is a distinct exclusion principle that cannot be reduced to either fermionic or bosonic statistics, representing a nontrivial extension of conventional quantum statistics. In Example $4$ the $R$-matrix is given by $\lambda_{ab}\epsilon_{cd}-\delta_{ac}\delta_{bd}$. The parameters $\lambda$ and $\epsilon$ are defined as special constant matrices. Their primary function is to enable the mathematical formulation of this exotic behaviour. The partition function in this example is given by $(1+{\mathscr{M}}x+x^2)  $. The system exhibits three distinct configurations: (1) A single configuration is permitted for the unoccupied state. (2) ${\mathscr{M}}$ distinct configurations are allowed for single-paraparticle occupation. (3) One specially constrained configuration is enabled for two-paraparticle occupation. Therefore the partition function is $(1+{\mathscr{M}}x+x^2)  $.  The system is allowed to have two paraparticles in the same mode, but only in one very specific way, determined by the R-matrix. This behaviour is fundamentally different from both: (1) Bosonic systems, where particles can freely occupy the same mode in multiple indistinguishable ways. (2) Fermionic systems, where no two particles can occupy the same mode at all due to the Pauli exclusion principle. For systems with ${\mathscr{M}} \geq 3$, the $R$-matrix becomes non-unitary, and the corresponding dynamics may no longer be Hermitian. This indicates a deviation from conventional quantum statistics\cite{wang2025particle}.

  The average occupation number is given by:
\[
\langle \tilde{n}_k \rangle = \frac{\sum_{n=0}^{n_{\max}} n z_R(x_k)}{z_R(x_k)}
\]

$\frac{d}{dx_k} z_R(x_k) = \frac{d}{dx_k} \left( \sum_{n=0}^{n_{\max}} d_n x_k^n \right) = \sum_{n=0}^{n_{\max}} n d_n x_k^{n-1}$. Multiplying both sides by $x_k$: $x_k \frac{d}{dx_k} z_R(x_k) = \sum_{n=0}^{n_{\max}} n d_n x_k^n$ which gives us:

\beq
\langle \tilde{n}_k \rangle  = \frac{x_k \frac{d}{dx_k} z_R(x_k)}{z_R(x_k)}
\eeq

\subsubsection{Case 1: Fermions}
Partition function of fermions is $(1+x)$, so the average occupation number is given by:

\beq
\langle \tilde{n} \rangle  = \frac{1}{ e^{\beta (\epsilon_k-\mu)} + 1}
\eeq

\subsubsection{Case 2: Bosons}
Partition function of fermions is ${(1-x)}^{-1}$, so the average occupation number is given by:

\beq
\langle \tilde{n} \rangle  = \frac{1}{ e^{\beta (\epsilon_k-\mu)} - 1}
\eeq

\subsubsection{Case 3: Taking Example 2 from Table 1}
Partition function of fermions is ${(1+x)}^{\mathscr{M}}$, so the average occupation number is given by:

\beq
 \langle \tilde{n} \rangle = \frac{{\mathscr{M}} e^{-\beta (\epsilon_k-\mu)} }{1 + e^{-\beta (\epsilon_k-\mu)} }
\eeq

\subsubsection{Case 4: Taking Example 3 from Table 1 (maximum 1 paraparticle per mode)}

One of the most interesting cases is that of Example $3$. The mode configurations are as follows: \( d_0 = 1 \) (vacuum state, no paraparticle), \( d_1 ={\mathscr{M}} \) (one paraparticle which can have any one of the \( {\mathscr{M}} \) flavours), and \( d_n = 0 \) for \( n \geq 2 \) (This means there can't be two paraparticles in one mode even if their flavours are different). For this specific case the partition function is given by $z_R(x) = 1 + {\mathscr{M}} x $, where $x= e^{-\beta (\epsilon_k-\mu)}$. The average occupation number is given by:

\beq
 \langle \tilde{n} \rangle = \frac{{\mathscr{M}}}{e^{\beta (\epsilon_k-\mu)} + {\mathscr{M}}} = \frac{1}{\frac{1}{{\mathscr{M}}} e^{\beta (\epsilon_k-\mu)} + 1}
\eeq

This differs from the case of fermions by the factor $\frac{1}{{\mathscr{M}}}$ multiplied to $e^{\beta (\epsilon_k-\mu)}$. 

\subsubsection{Case 5: Taking Example 4 from Table 1 (maximum 2 paraparticles per mode)} 

Partition function is given by: $ z_R(x) = 1 + {\mathscr{M}} x + x^2 $

\beq
\langle \tilde{n} \rangle = \frac{{\mathscr{M}} x + 2 x^2}{1 + {\mathscr{M}} x + x^2}
\eeq

\begin{figure}
  \centering
  \includegraphics[width=0.5\textwidth]{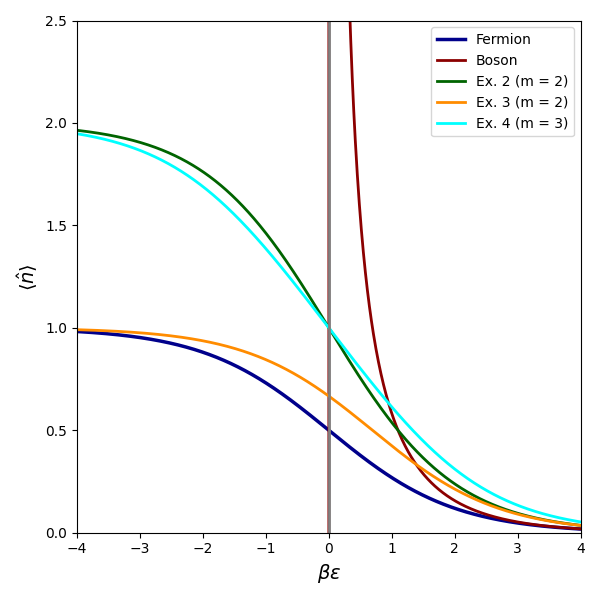}  
  \caption{Graph of average occupation number vs. $\beta \epsilon$ for particles with various partition functions. This agrees with Figure 1 in reference\cite{wang2025particle}.}
  \label{fig:myplot}
\end{figure}


\section{Thermodynamics of the Paraparticles of Example 3}

The partition function of mode $k$ is:

\beq
z_k = 1 + {\mathscr{M}} e^{-\beta(\epsilon_k - \mu)}
\eeq

The Thermodynamic Potential (analogue of the Gibbs free energy) for $k^{th}$ mode is:

\beq
\Omega_k = -k_B T \ln(1 + {\mathscr{M}} e^{-\beta(\epsilon_k - \mu)}).
\eeq



Since $\Omega_k = -k_B T \ln z_k$, one immeditely recovers $\langle \tilde{n_k} \rangle =-\frac{\partial \Omega_k}{\partial \mu} = \frac{1}{\frac{1}{{\mathscr{M}}} e^{\beta(\epsilon_k - \mu)}+1} $ consistent with Eq. (22).


\subsection*{Thermodynamic Potential of the whole gas:}

\beq
\Omega = \sum_k \Omega_k =  -k_B T \sum_k ln(1 + {\mathscr{M}} e^{-\beta(\epsilon_k - \mu)})
\eeq

\subsection*{Entropy of paraparticles:}
The entropy is given by \cite{pathria}:
\beq
S = -\left.\frac{\partial\Omega}{\partial T}\right|_{V,\mu} 
\eeq

\beq
S = k_B \sum_k \left( ln(1 + {\mathscr{M}} e^{-\beta(\epsilon_k - \mu)}) + \frac{{\mathscr{M}} e^{-\beta(\epsilon_k - \mu)  }\beta (\epsilon_k - \mu) }{1 + {\mathscr{M}}e^{-\beta(\epsilon_k - \mu)}} \right) 
\eeq
Let us redefine $\langle \tilde{n}_k \rangle = n_k$ for simplicity. Now,
\[
 n_k = \frac{1}{{\frac{1}{\mathscr{M}}} e^{\beta(\epsilon_k - \mu)} + 1}
\]

or
\beq
 e^{-\beta(\epsilon_k - \mu)}  = \frac{1}{M} \frac{n_k}{1 - n_k}
\eeq

\[
S = k_B \sum_k \left( ln\left(1 + \frac{n_k}{1 - n_k}\right) + \beta(\epsilon_k - \mu) n_k\right),
\]

we have

\[
\beta(\epsilon_k - \mu) = \ln M + \ln\left(\frac{1-n_k}{n_k}\right).
\]
Then
\[
S = k_B \sum_k \left( \ln\left(\frac{1}{1-n_k}\right) + \left[\ln M + \ln\left(\frac{1-n_k}{n_k}\right)\right] \frac{n_k}{1} \right).
\]

\beq
S = -k_B \sum_k \left( n_k \ln n_k + (1-n_k) \ln (1-n_k) - n_k \ln {\mathscr{M}} \right)
\eeq


Heat capacity at constant volume and chemical potential $(\mu)$ is given by:

\beq
C_V = T \left.\frac{\partial S}{\partial T}\right|_{\mu, V} = -\beta \left.\frac{\partial S}{\partial \beta}\right|_{\mu, V} 
\eeq

\beq
\frac{dn_k}{d\beta} = -\frac{\frac{1}{M} e^{\beta(\epsilon_k - \mu)} (\epsilon_k - \mu)}{\left(\frac{1}{M} e^{\beta(\epsilon_k - \mu)} + 1\right)^2} 
\eeq

From Equation (31):


\begin{multline}
C_V = -\beta \frac{\partial S}{\partial \beta} = k_B \beta \sum_k ( \frac{\partial n_k}{\partial \beta} \ln n_k + \frac{n_k}{n_k} \frac{\partial n_k}{\partial \beta} \\
 - \frac{\partial n_k}{\partial \beta} \ln (1 - n_k) + ( \frac{1-n_k}{1-n_k} ) ( \frac{\partial n_k}{\partial \beta} ) \\
- \frac{\partial n_k}{\partial \beta} \ln {\mathscr{M}} )
\end{multline}


\[
= \frac{1}{T} \sum_k \frac{\partial n_k}{\partial \beta} \left[ ln\left(\frac{n_k}{1-n_k}\right) - ln{\mathscr{M}} \right]
\]
We had
\[
 \left( \frac{n_k}{1-n_k} \right) = {\mathscr{M}} e^{-\beta(\epsilon_k - \mu)}
\]

\[
C_V = \frac{1}{T} \sum_k \frac{\partial n_k}{\partial \beta} \left[ ln{\mathscr{M}} - \beta(\epsilon_k - \mu) - ln{\mathscr{M}} \right]
\]

\beq
C_v = -\frac{\beta}{T} \sum_k \frac{\partial n_k}{\partial \beta}(\epsilon_k - \mu)
\eeq


From equation (33):

\[
\frac{\partial n_k}{\partial\beta} = -(\epsilon_k - \mu) \left( \frac{1}{\frac{1}{{\mathscr{M}}} e^{\beta (\epsilon_k - \mu)}+ 1}  \right) \left( \frac{ \frac{1}{{\mathscr{M}}} e^{\beta (\epsilon_k - \mu)} }{\frac{1}{{\mathscr{M}}} e^{\beta (\epsilon_k - \mu)}+ 1} \right)
\]

\beq
= -(\epsilon_k - \mu) n_k (1 - n_k) 
\eeq

Therefore

\[
C_V = \frac{\beta}{T} \sum_k (\epsilon_k - \mu)^2 n_k (1 - n_k)
\]

\beq
C_V = k_B \beta^2 \sum_k (\epsilon_k - \mu)^2 n_k (1 - n_k) 
\eeq


Introducing the paraparticle density of states (refer to Appendix A):

\beq
C_V = k_B \beta^2 \int_0^\infty d\epsilon  \, \rho(\epsilon) \, (\epsilon - \mu)^2 n(\epsilon)(1 - n(\epsilon)) 
\eeq

Let us study the behaviour in the low temperature limit.

$k_B T \ll \mu$  the factor   $n(\epsilon) {(1 - n(\epsilon))}$  acts like a delta function.

\beq
C_V =k_B \, \rho(\mu) \int_0^\infty d\epsilon \left[ \beta (\epsilon - \mu) \right]^2 n(\epsilon)(1 - n(\epsilon))
\eeq

Let us evaluate the integral and substitute:


\[
x = \beta (\epsilon - \mu), \quad dx = \beta \, d\epsilon 
\quad \text{or} \quad d\epsilon = \frac{1}{\beta} dx
\]

\[
C_V = \frac{k_B \, \rho(\mu)}{\beta} 
\int_{-\beta \mu}^{\infty} dx \, x^2 \, 
\frac{\tfrac{1}{{\mathscr{M}}} e^{\beta(\epsilon - \mu)}}{\left(\tfrac{1}{{\mathscr{M}}} e^{\beta(\epsilon - \mu)} + 1\right)^2}
\]

\beq
= k_B^2 \, T \, \rho(\mu) 
\int_{-\infty}^{\infty} dx \, x^2 \,
\frac{\tfrac{1}{{\mathscr{M}}} e^x}{\left(\tfrac{1}{{\mathscr{M}}} e^x + 1\right)^2}
\eeq


Consider the integral $I_1$

\[
I_1 = \int_{0}^{\infty} dx \, x^2 \, 
\frac{\tfrac{1}{{\mathscr{M}}} e^x}{\left(\tfrac{1}{{\mathscr{M}}} e^x + 1\right)^2}
\]

\[
= \int_{0}^{\infty} dx \, x^2 \,
\frac{1}{\tfrac{1}{{\mathscr{M}}} e^x \left( 1 + {\mathscr{M}} e^{-x}\right)^2}
\]

\[
= M \int_{0}^{\infty} dx \, x^2 \, e^{-x} \,
\left( \frac{1}{1 + {\mathscr{M}} e^{-x}} \right)
\left( \frac{1}{1 + {\mathscr{M}} e^{-x}} \right)
\]

\[
= M \int_{0}^{\infty} dx \, x^2 \, e^{-x} \,
\left( \frac{1}{1 + e^{-(x - \ln {\mathscr{M}})}} \right)
\left( \frac{1}{1 + e^{-(x - \ln {\mathscr{M}})}} \right)
\]


Let
\[
x - \ln {\mathscr{M}} = y, 
\quad x = y + \ln {\mathscr{M}}, 
\quad dx = dy
\]

\[
I_1 = {\mathscr{M}} \int_{-\ln {\mathscr{M}}}^{\infty} dy \, (y + \ln {\mathscr{M}})^2 \, e^{-y} 
\left( \frac{1}{1 + e^{-y}} \right)
\left( \frac{1}{1 + e^{-y}} \right)
\]
The expression in each parenthesis is the sum of a geometric progression. In total, we have,
\[
= \int_{-\ln {\mathscr{{\mathscr{M}}}}}^{\infty} dy \, (y + \ln M)^2 \, e^{-y} 
\left( 1 - 2 e^{-y} + 3 e^{-2y} - 4 e^{-3y} + \cdots \right)
\]

Going back to \(x\):

\[
e^{-y} = e^{-(x - \ln {\mathscr{M}})} = e^{-x} \, e^{\ln {\mathscr{M}}} ={\mathscr{M}} e^{-x}
\]

Therefore,

\begin{multline}
I_1 = \int_{0}^{\infty} dx \, x^2 \, e^{-x} \, {\mathscr{M}} \\
\left( 1 - 2 e^{-x} ({\mathscr{M}}) + 3 e^{-2x} ({\mathscr{M}})^2 - 4 e^{-3x} ({\mathscr{M}})^3 + \cdots \right)
\end{multline}


\begin{multline}
= {\mathscr{M}} \int_{0}^{\infty} dx \, x^2 e^{-x} \\
\left( 1 - 2{\mathscr{M}} e^{-x} + 3{\mathscr{M}}^2 e^{-2x} - 4{\mathscr{M}}^3 e^{-3x} + \cdots \right)
\end{multline}

We use the standard integral
\[
\int_{0}^{\infty} dx \, x^2 e^{-n x} = \frac{2}{n^3}.
\]

Therefore,

\[
I_1= 2{\mathscr{M}} \left( 1 - {\mathscr{M}} \frac{1}{2^2} + {\mathscr{M}}^2 \frac{1}{3^2} - {\mathscr{M}}^3 \frac{1}{4^2} + \cdots \right)
\]

\[
= -2 Li_2(-{\mathscr{M}}).
\]
Here $Li_2$ is the dilogarithmic function. In general $Li_{s}(z)=\sum_{k=1}^\infty \frac{z^k}{k^s}, \quad  |z|<1$.

Refer back to equation $(40)$ and define 
\bn
I_2 = \int^{0}_{-\infty} dx \, x^2 \, \frac{\frac{1}{{\mathscr{M}}}e^x}{\frac{1}{{\mathscr{M}}}e^x + 1}
\en

\[
I_2 = \int_{0}^{\infty} dx \, x^2 \, \frac{\frac{1}{{\mathscr{M}}}e^{-x}}{\frac{1}{{\mathscr{M}}}e^{-x} + 1}
\]
\[
= \frac{1}{{\mathscr{M}}} \int_{0}^{\infty}  dx \, x^2 \, e^{-x} \frac{1}{(1+ e^{-x-ln{\mathscr{M}}})(1+ e^{-x-ln{\mathscr{M}}})}
\]

Let 

\[
y = x + ln{\mathscr{M}} \quad \text{or} \quad x = y -{\mathscr{M}}
\]

\[
I_2 = \frac{1}{{\mathscr{M}}} \int_{ln {\mathscr{M}}}^{\infty} dy {(y-ln{\mathscr{M}})}^2 \, e^{-y+ln{\mathscr{M}}} \left( \frac{1}{1+e^{-y}} \frac{1}{1+e^{-y}}\right)
\]

\[
I_2 = \frac{1}{{\mathscr{M}}} \int_{ln {\mathscr{M}}}^{\infty} dy {(y-ln{\mathscr{M}})}^2 \, e^{-y} \left(1 - 2e^{-y} + 3e^{-2y}- 4 e^{-3y} + \dots\right)
\]
Going back to x:
\beq
I_2 = \frac{1}{{\mathscr{M}}} \int_{0}^{\infty} dx \, x^2 \, e^{-x} \left(1 - 2\frac{e^{-x}}{{\mathscr{M}}} + 3\frac{e^{-2x}}{{\mathscr{M}}^2}- 4 \frac{e^{-3x}}{{\mathscr{M}}^3} + \dots\right)
\eeq
It is exactly the same expression as in equation $(42)$ except ${\mathscr{M}}$ going to $\frac{1}{{\mathscr{M}}}$. Therefore, the final expression of the heat capacity is as follows:

\beq
C_V = 2 {k_B}^2 T \rho(\mu) \left(-Li_2 (- {\mathscr{M}}) -Li_2 \left(- \frac{1}{{\mathscr{M}}}\right)\right)
\eeq
Let us check whether this expression goes to the standard expression for electrons for ${\mathscr{M}}=1$, we have:

\[
-Li_2 (-1)=\frac{\pi^2}{12}
\]
Therefore,
\beq
C_V = \frac{\pi^2}{3} {k_B}^2 T \rho(\mu) .
\eeq

We notice that in the case of ${\mathscr{M}}=1$ we do get back the standard case of electrons, as it should. Let us compare the ratio of the heat capacity of paraparticles to that of the electrons: 
\beq
\frac{{C_V}^{paraparticles}}{{C_V}^{electrons}} = 1 + \frac{1}{2} \frac{{(ln {\mathscr{M}}})^2}{\zeta(2)}.
\eeq

Here $\zeta$ is the Riemann Zeta function. The above equation is our final result. It provides a handle on the thermodynamic detection of the presence of paraparticles, by just doing the careful heat capacity measurements in appropriate systems\cite{wang2025particle}. Similar expression can be derived for Example $4$ in Table $1$, but it turns out to be very cumbersome.

\section{Conclusion}
We have reviewed the basic foundations regarding paraparticles laid in Ref.\cite{wang2025particle}. We extended that work by computing various thermodynamical quantities for paraparticles including entropy. We also computed heat capacity of paraparticles for the most interesting example, that is, Example $3$ in Table $1$. We compared that expression with that of the standard expression for fermions. We noticed that there is an excess heat capacity in the case of paraparticles which is proportional to the square of the logarithm of the flavour index ${\mathscr{M}}$ of the paraparticles. We propose that our expression can provide a foundation for the experimental detection of paraparticles.

\section{Acknowledgement} 
The authors are thankful to Zhiyuan Wang for the correspondence and the very useful
comments. We also thank Borivoje Daki\'{c} for bringing the related work of
Medina S\'anchez and Daki\'{c}~\cite{MedinaDakic2024} to our attention and
for helpful correspondence clarifying its connection to the thermodynamic
results presented here.

\appendix
\section{Paraparticle Density of States}

Let us define paraparticle density of states in 3D as $\rho(\epsilon)$.
Energy of paraparticle in mode $k$ is given by the dispersion relation $\epsilon_k = \frac{\hbar_k}{2 m}$\cite{wang2025personal}. The number of states in a spherical shell of radius $k$ and thickness $dk$ is given by:
\[
dN ={\mathscr{G}} \frac{V}{({2\pi})^3} 4 \pi k^2 dk  = {\mathscr{G}} \frac{V}{2\pi^2} k^2 dk 
\]

where V represents the total volume of the system and ${\mathscr{G}}$ is the degeneracy that defines how many paraparticles are allowed to occupy a mode (for example 3  ${\mathscr{G}}=1$ ). The density of states in $k$-space is given by $\rho(k)=\frac{dN}{dk}$. On substituting, it gives:
\[
\rho(k) = {\mathscr{G}} \frac{ V k^2}{2\pi^2}
\]

The total number of particles at $T = 0$ can then be computed as:
\[
N = \int_0^{k_{GW}} \rho(k)\, dk = \frac{{\mathscr{G}}V}{2\pi^2}  \frac{k_{GW}^3}{3}
\]

Here $k_{GW}$ is a general wavevector which generalises the definition of the Fermi wavevector. Solving for the general wavevector $k_{GW}$ gives:
\[
k_{GW} = \left( \frac{6\pi^2 N}{{\mathscr{G}}V} \right)^\frac{1}{3}
\]

To compute the density of states in terms of energy, use the relation \cite{wang2025personal}:
\[
\epsilon = \frac{\hbar^2 k^2}{2m} 
\]

Thus, $\rho(k)\,dk = {\mathscr{G}} \frac{V}{2\pi^2} \cdot k^2\, dk
= \frac{{\mathscr{G}}V}{2\pi^2} \left( \frac{2m\epsilon}{\hbar^2} \right) \left( \frac{m}{\hbar^2} \right) \left( \frac{\hbar^2}{2m\epsilon} \right)^{1/2} d\epsilon$. Simplifying: $\rho(\epsilon) ={\mathscr{G}} \frac{Vm}{2\pi^2 \hbar^2} \left( \frac{2m}{\hbar^2} \right)^{1/2} \epsilon^{1/2}$

Which can also be written as 
\beq
\rho(\epsilon) = {\mathscr{G}}\frac{V}{4\pi^2} \left( \frac{2m}{\hbar^2} \right)^\frac{3}{2} \sqrt{\epsilon}= {\mathscr{G}} A \sqrt{\epsilon}
\eeq
here, 
\[
A= \frac{V}{4\pi^2} \left( \frac{2m}{\hbar^2} \right)^\frac{3}{2}
\]
where $m$ is the mass of the paraparticle.


\begin{thebibliography}{201}
\bibitem{wang2025personal}
Zhiyuan Wang, 
``Private communication. If there is no interaction between paraparticles, then the many-body problem reduces to a one-particle problem, and the resulting single-particle Schrödinger’s equation is the same for a single fermion or a single boson, because statistics can not show up in a single-particle problem. Thus, a free paraparticle will obey Schrödinger’s equation.''
\end{thebibliography}
\end{document}